\documentstyle[prl,aps,preprint,amssymb]{revtex}
\begin{document}
\draft
\title{Disorder Induced Diffusive Transport In Ratchets}
\author{ M.N. Popescu$^1$, C.M. Arizmendi$^{1,2}$, A. L.
Salas- Brito$^{1,3}$, and F. Family$^1$}
\address {$^1$Department of Physics, Emory University,
Atlanta, GA 30322,  USA\\
$^2$Depto. de F\'{\i}sica, Facultad de Ingenier\'{\i}a,
Universidad Nacional de Mar del Plata,\\  Av. J.B. Justo
4302, 7600 Mar del Plata, Argentina\\
$^3$Laboratorio de Sistemas Din\'amicos, Departamento de
Ciencias B\'asicas,\\ Universidad Aut\'onoma
Metropolitana-Azcapotzalco, Apartado Postal 21-726,\\ 
Coyoac\'an 04000 D.\ F., M\'exico}
\date{\today}
\maketitle

\begin{abstract} 
The effects of quenched disorder on the overdamped motion 
of a driven particle on a periodic, asymmetric potential is 
studied. While for the unperturbed potential the transport 
is due to a regular drift, the quenched disorder induces a 
significant additional chaotic ``diffusive'' motion. 
The spatio-temporal evolution of the statistical ensemble is 
well described by a Gaussian distribution, implying a chaotic 
transport in the presence of quenched disorder.
\end{abstract}

Stochastic models known as {\it thermal ratchets} or 
{\it correlation ratchets} \cite{general}, in which a non-zero 
net drift velocity may be obtained from time correlated 
fluctuations interacting with broken symmetry structures 
\cite{magnasco}, have recently received much attention. This
interest is due to the wide range of possible applications
of these models for understanding such systems as molecular 
motors \cite{motors}, nano-scale friction \cite{friction}, 
surface smoothening \cite{barabasi}, coupled Josephson 
junctions \cite{josephson}, optical ratchets and directed motion 
of laser cooled atoms \cite{optics}, mass separation and 
trapping schemes at the microscale \cite{separation}. Recently, 
spatial disorder in thermal ratchets has been shown to reduce 
the characteristic drift speed \cite{Marchesoni,Harms}. 
Little is known, however, about the effects of quenched 
spatial disorder on the regular or diffusive motion in 
otherwise periodic potentials 
\cite{Marchesoni,Harms,Garcia,Radons,Igloi}. 

Diffusion-like motion is observed in many types of
deterministic systems. In particular, it has been shown that
in  deterministic chaotic systems, diffusion can be normal
\cite{normal}, with the mean-square displacement
$\langle x^2 \rangle$ proportional to time $t$ ($\langle x^2
\rangle \sim t$), or it can be anomalous \cite{anomalous}, 
with $\langle x^2 \rangle \sim t^\gamma$, (enhanced for 
$\gamma > 2$, dispersive for $1 < \gamma < 2$), or have a 
logarithmic time dependence ($\gamma = 0$) \cite{log}.

In the present work we report on an unusual behavior 
that  occurs in the case of an overdamped ratchet subject to 
an external oscillatory  drive: quenched disorder induces a 
normal diffusive transport in addition to the drift due to the 
external drive. For the parameter range considered, this 
process is observed even for very small perturbations. 
Moreover, this diffusive  motion is enhanced by higher values 
of the quenched disorder. In fact, for high enough disorder 
the diffusive motion is of the same order of magnitude as the 
regular drift. The possibility of having large fluctuations, 
of the same order of magnitude as the average  velocity, can  
be of great importance for a correct interpretation of 
experimental  results. This may be of particular importance 
in studies of friction, in understanding the motion of 
nanoclusters or monolayers sliding on surfaces, as well as 
for designing particles separation techniques.

We consider the one-dimensional, overdamped motion of a
particle (in dimensionless  units) on a disordered ratchet:
\begin{equation}
\gamma {{d x} \over {d t}} = \cos (x) + \mu \cos (2x) + \Gamma
\sin(\omega t) + \alpha~\xi(x). 
\label{motion} 
\end{equation}
Here, $\gamma$ is the damping coefficient, $\Gamma$ and
$\omega$ are, respectively, the amplitude and frequency of
an external oscillatory forcing, and $\alpha~\xi(x)$ is the
force due to the quenched disorder. For the present study,
$\xi(x) \in [-1,1]$ are independent, uniformly distributed
random variables with no spatial correlations,  and $\alpha
\geq 0$ is the amount of quenched disorder. This corresponds
to a  piecewise constant force on the interval 
$\biglb[ 2 k \pi, 2 (k + 1) \pi  \bigrb)$, 
$k \in \Bbb Z$. The unperturbed ratchet potential, 
\begin{equation} 
U(x) = - \sin(x) - \mu \sin(2x)
\label{potential} 
\end{equation} 
has been the subject of extensive recent studies
\cite{Bartussek,Jung,Lindner}. The quenched disorder term
($\alpha \neq 0$) is expected to give either a more  realistic
representation of a real substrate or potential landscape, 
or to model fluctuations in DC current amplitude, as for arrays 
of Josephson junctions.

It is well known \cite{Bartussek,Jung} that in the absence
of  quenched disorder ($\alpha = 0$) there are unbounded
solutions of Eq.\ (\ref{motion}), provided that the driving
amplitude $\Gamma$ is  large enough. These solutions tend
asymptotically to a constant average  velocity independent of
the initial conditions \cite {Jung}. We have  identified a
set of parameters where, in the absence of disorder, the
system  shows non-zero current (regular transport).
Specifically, we have  selected $\gamma = 1.0$, $\mu=0.25$,
$\omega = 0.1$, and we have  studied the behavior for several
values of $\Gamma$ with $\Gamma \geq 1.4$ (see below).

For $\alpha > 0$ the periodicity of the unperturbed potential
is  destroyed as a result of the spatial randomness, and 
solutions  of Eq.\ (\ref{motion}) begin to show a very
complex behavior, including chaotic motion. The chaotic behavior
is characterized by the rate of divergence of trajectories
starting from very close initial conditions, in other words by
the leading positive Lyapunov exponent. For a given, fixed 
realization of the quenched disorder, and for several values of
$\Gamma \geq 1.4$, we have computed Lyapunov exponents over 
trajectories starting from origin. We have found positive 
Lyapunov exponents $\Lambda$ ranging from 2.55, for 
$\Gamma = 1.4$, to 3.22 for $\Gamma = 1.76$, which shows a 
very strong chaotic behavior. As a consequence of this chaotic 
behavior, the motion of the particle in the perturbed 
potential should be characterized by ensemble averages 
performed not only over realizations  of disorder, but also 
over the spatial distribution of the positions of the 
particle in a given realization of the quenched disorder. 

Numerical solutions of Eq.\ (\ref{motion}) were obtained
using a  variable step Runge-Kutta-Fehlberg method
\cite{NumRec}. Averages  were performed over ensembles of
5000 trajectories starting from different initial conditions 
very close to the origin $x = 0$. The ensemble described above 
was left to evolve  for 1000 external drive periods $T$, and 
every 10 periods the positions $x(t)$ were stored for further 
analysis.

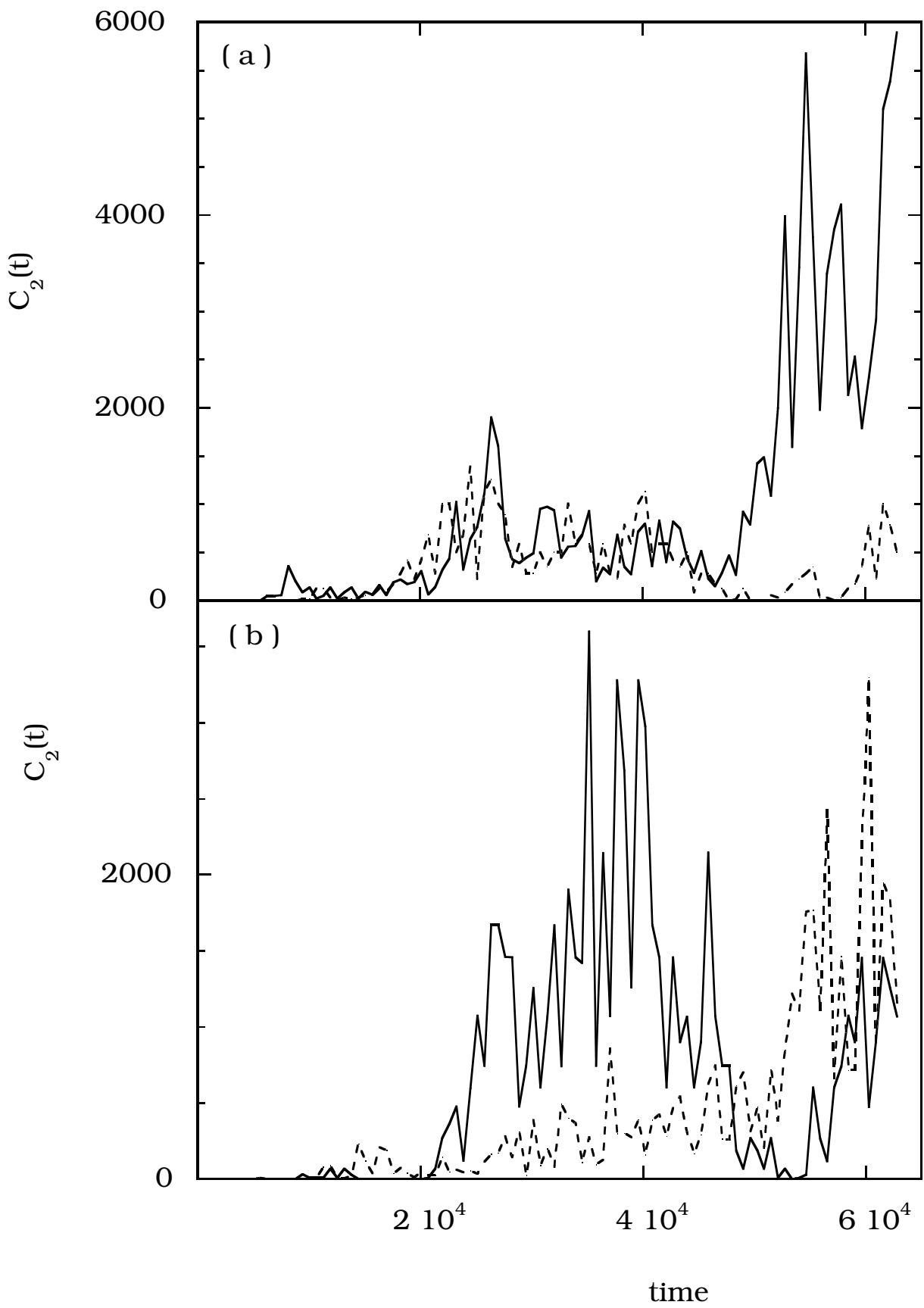
\begin{figure}
\caption{Second moment $C_2(t)$ as a function of time for two
different,  fixed realizations of quenched disorder (solid,
respectively dashed lines).  The parameters used in Eq.\
(\ref{motion}) are $\gamma = 1$, $\mu=0.25$, 
$\Gamma = 1.76$, $\omega = 0.1$, and $\alpha = 0.05$ (panel
(a)),  respectively $\alpha = 0.10$ (panel (b)).}
\label{fig1}
\end{figure}

We have first analyzed the motion in a given realization of 
disorder. In Fig.\ \ref{fig1} we show results for the second 
moment, 
$C_2(t) = \langle {(x(t)-\langle x(t) \rangle)}^2 \rangle$, 
where $\langle~\dots~\rangle$ means average over the
ensemble, as a function of the time $t$, for two different,
fixed realizations  of quenched disorder, and for two 
values of the disorder parameter $\alpha$, $\alpha = 0.05$ 
(panel (a)), respectively $\alpha = 0.10$ (panel (b)).  
The most striking feature is the fact that the second moment 
which is zero in the absence of disorder (corresponding to a 
purely deterministic motion), becomes non-zero in the presence 
of the perturbation. The non-zero second moment confirms the
chaotic behavior mentioned above, showing a {\it
disorder-induced} sensitive dependence on the initial 
conditions. It can be seen also that the time-dependence of
the second  moment is very complicated, and it is dependent
on both the realization  of quenched disorder and the
amplitude $\alpha$ of disorder. 

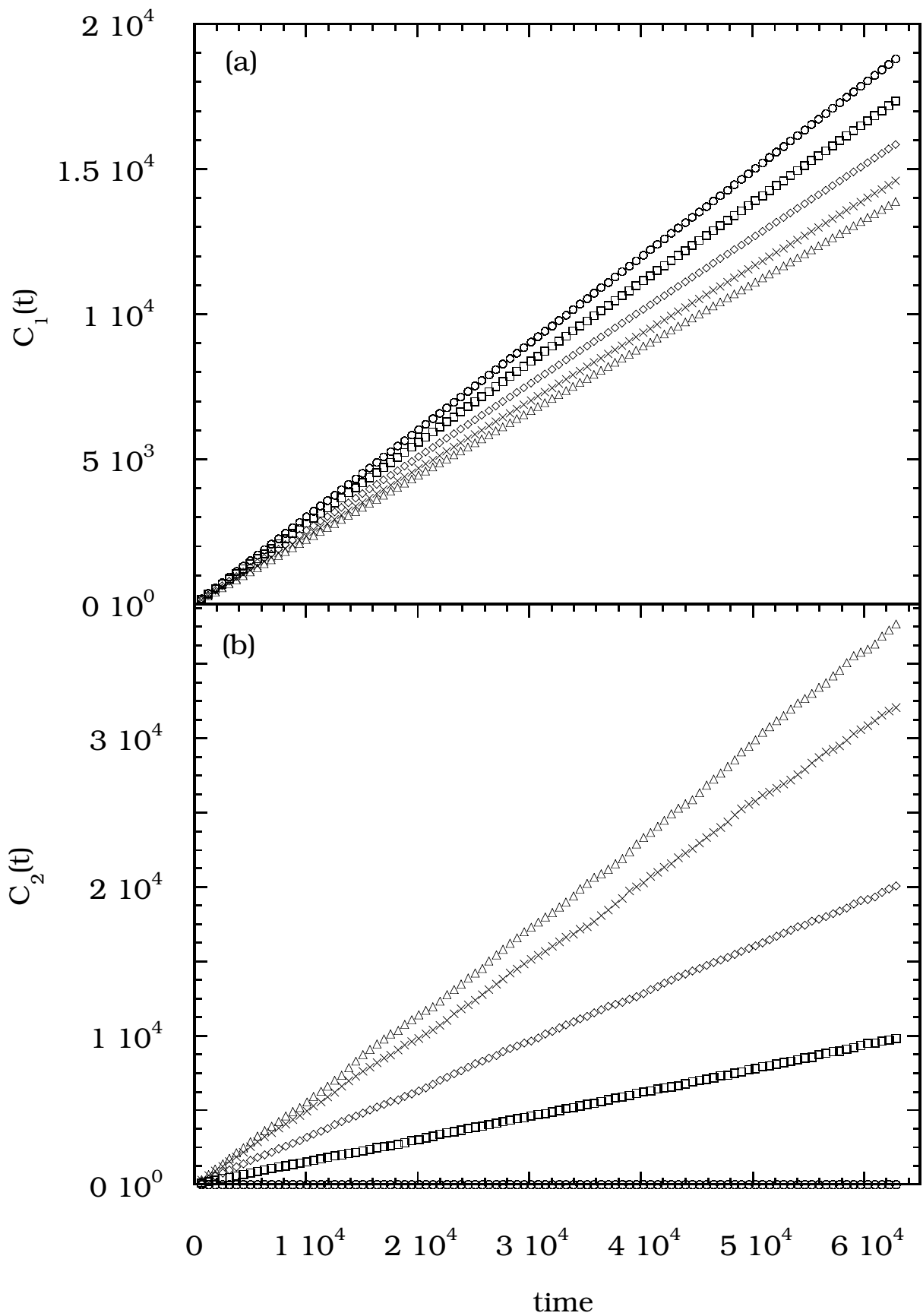
\begin{figure}
\caption{(a) First moment $C_1(t)$ as a function of time for
several  values of the amount of quenched disorder parameter
$\alpha$. From top to  bottom, $\alpha =
0,~0.05,~0.10,~0.15,~0.20$.  (b) Second moment $C_2(t)$ as a
function of time for several  values of the amount of
quenched disorder parameter $\alpha$. From bottom to  top,
$\alpha = 0,~0.05,~0.10,~0.15,~0.20$.  The parameters used in
Eq.\ (\ref{motion}) are $\gamma = 1$, $\mu=0.25$, 
$\Gamma = 1.50$, and $\omega = 0.1$.}
\label{fig2}
\end{figure}

In order to perform averages over the realizations of
disorder, we have used for each trajectory a different
random sequence $\xi(x)$. In this way, the averages over the
ensemble of trajectories include also  averages over
realizations of disorder. Figure\ \ref{fig2} shows results 
for the first two moments, $C_1(t) = \langle x(t) \rangle$
and 
$C_2(t) = \langle {(x(t)-\langle x(t) \rangle)}^2 \rangle$,
as a function  of the time $t$ for $\Gamma=1.5$ for several
values of disorder  parameter $\alpha$. In contrast to
averages over a given "landscape",  in this case both first
and second moment show an asymptotic linear  dependence on
time $t$, $C_1(t) \simeq v(\alpha)~t$, 
$C_2(t) \simeq D(\alpha)~t$. We have considered several other 
values $1.40 \leq \Gamma \leq 1.76$, and we have observed the
same linear  behavior for all $\alpha$ values below a
threshold value which depends  on $\Gamma$. The quenched
disorder induces fluctuations in the spatial position 
around the average value in our system, and the dynamics is 
no longer regular, but rather consists in a superposition of 
regular drift and diffusion-like chaotic motion. Moreover, 
even for reasonably small amounts of disorder, for example 
$\alpha = 0.1$, it can be seen that these spatial 
fluctuations are of the same order of magnitude as the first 
moment, so the knowledge of a particular $x(t)$ no longer 
gives relevant information about the position of the center 
of mass of the distribution, an observation that can be of 
importance in studies of friction, particularly the sliding 
motion of clusters on surfaces \cite{friction}.

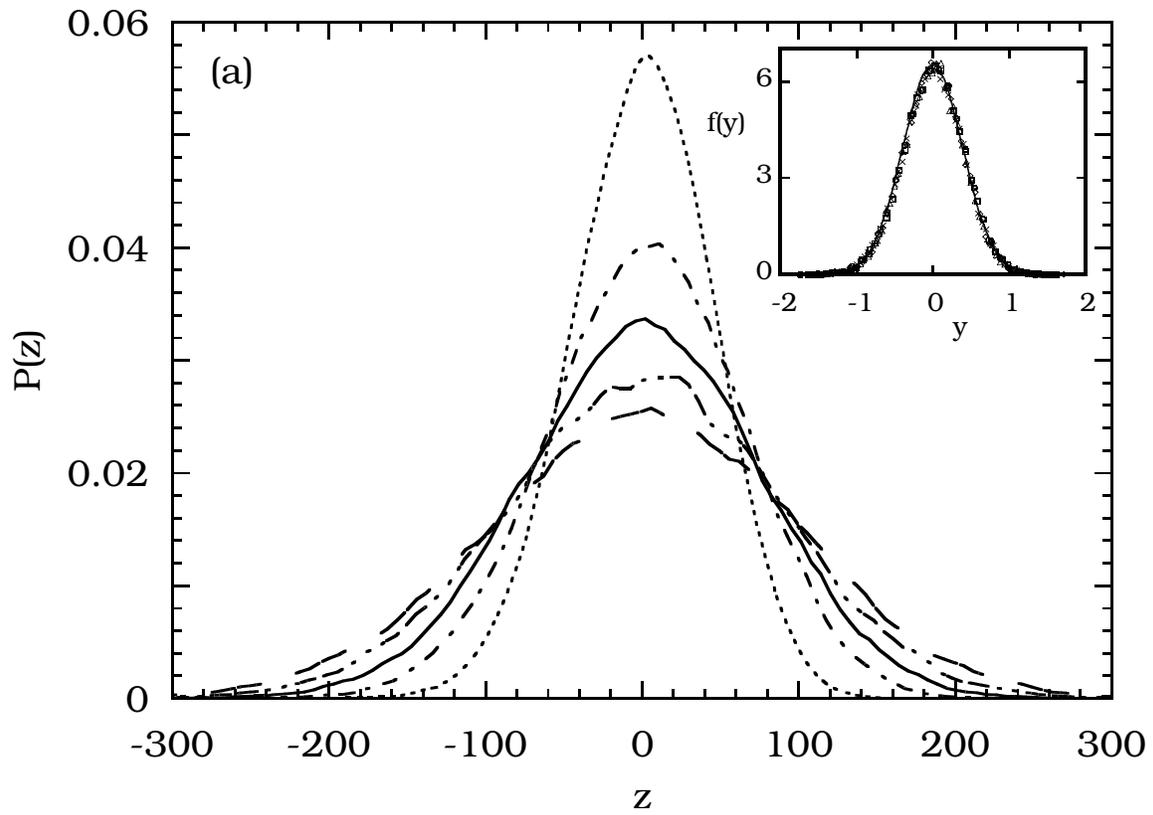
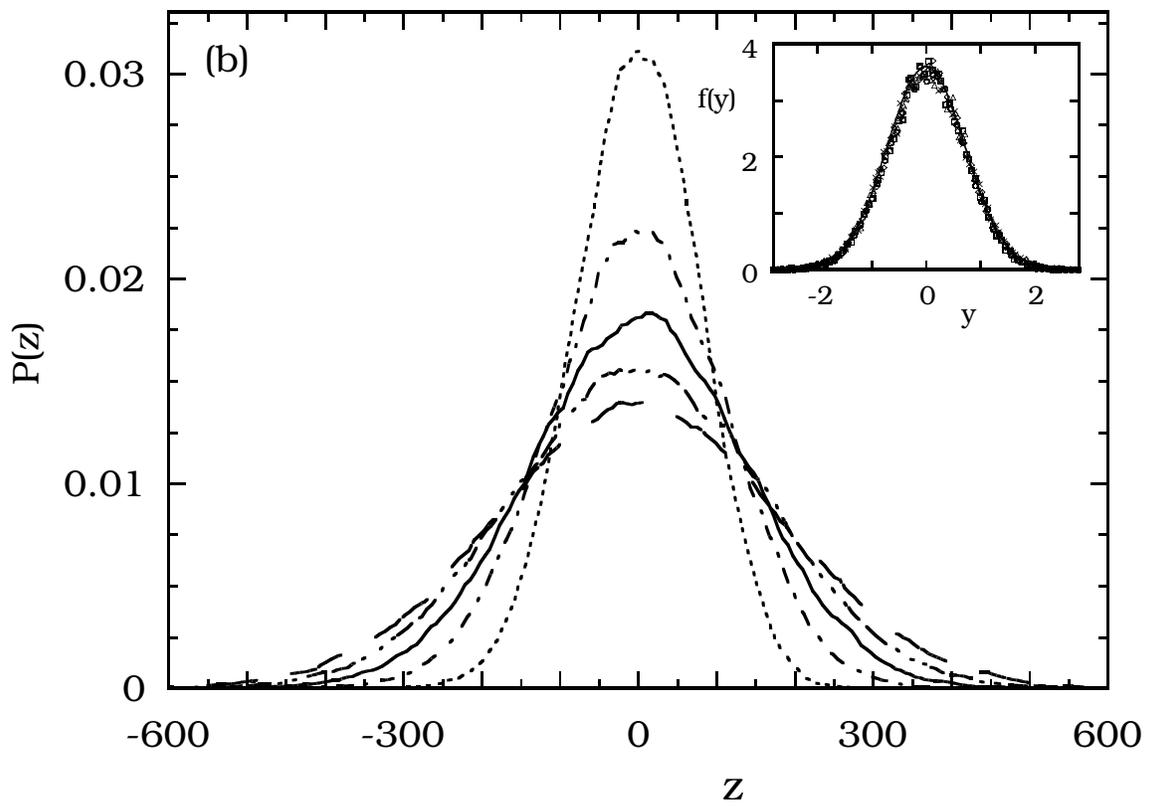
\begin{figure}
\caption{Time evolution of the unscaled spatial distributions 
$P(z)$, where $z = x-\langle x \rangle$; from top to bottom, 
time $t = 20~T, 40~T,\dots, 100~T$. The inset shows the scaled distributions $P(z) \times \sqrt{t}$ {\it vs} $z/\sqrt{t}$. 
The solid line in the insets  shows the theoretical, asymptotic 
Gaussian form. The parameters  used in Eq.\ (\ref{motion}) are 
$\gamma = 1$, $\mu=0.25$, $\Gamma = 1.50$, $\omega = 0.1$, and 
$\alpha = 0.05$ (panel (a)), respectively $\alpha = 0.10$ 
(panel (b)).}
\label{fig3}

\end{figure}

The fluctuations in the position are characterics of chaotic 
behavior in deterministic systems. The description of the 
initial ensemble is then given  by a probability distribution 
function $p_t(x)$, whose first two moments are linear in time 
as we have shown above. We have also calculated the higher 
order cumulants $C_n(t)$, for $n \leq 6$, and we have found 
that they increase  slower than $t^{n/2}$. Therefore, $p_t(x)$ 
is asymptotically a Gaussian, and it is determined by the first 
two moments \cite{Jung}. In Figure\ \ref{fig3} we show the 
distributions $P(z)$, where $z = x-\langle x \rangle$, for 
$\Gamma=1.50$ and two values of disorder parameter 
$\alpha = 0.05$ (panel (a)), and $\alpha = 0.10$ (panel (b)), 
at several  times $t$, and the scaled distributions 
$f(y) = P(z) \times \sqrt(t)$, where $y = z/\sqrt {t}$; one 
can see that the distribution is indeed well  aproximated by a 
Gaussian. This asymptotic Gaussian behavior also supports  the conclusion that the motion is chaotic, as it was shown 
by Jung {\it et al} \cite{Jung}. The reason for this chaotic 
behavior is the existence of discontinuities in the velocity at 
$x_k = 2 \pi k$, where $k$ is an integer, introduced by the 
quenched disorder perturbation. These random kicks keep into a transitory regime the trajectories that in the absence of 
disorder would have asymptotically converged to the asymptotic 
constant speed state mentioned above. This ``mixing'' of 
transitory regimes causes the chaotic behavior, and we emphasize 
again  that it is an effect due solely to the perturbation 
induced by quenched spatial disorder.

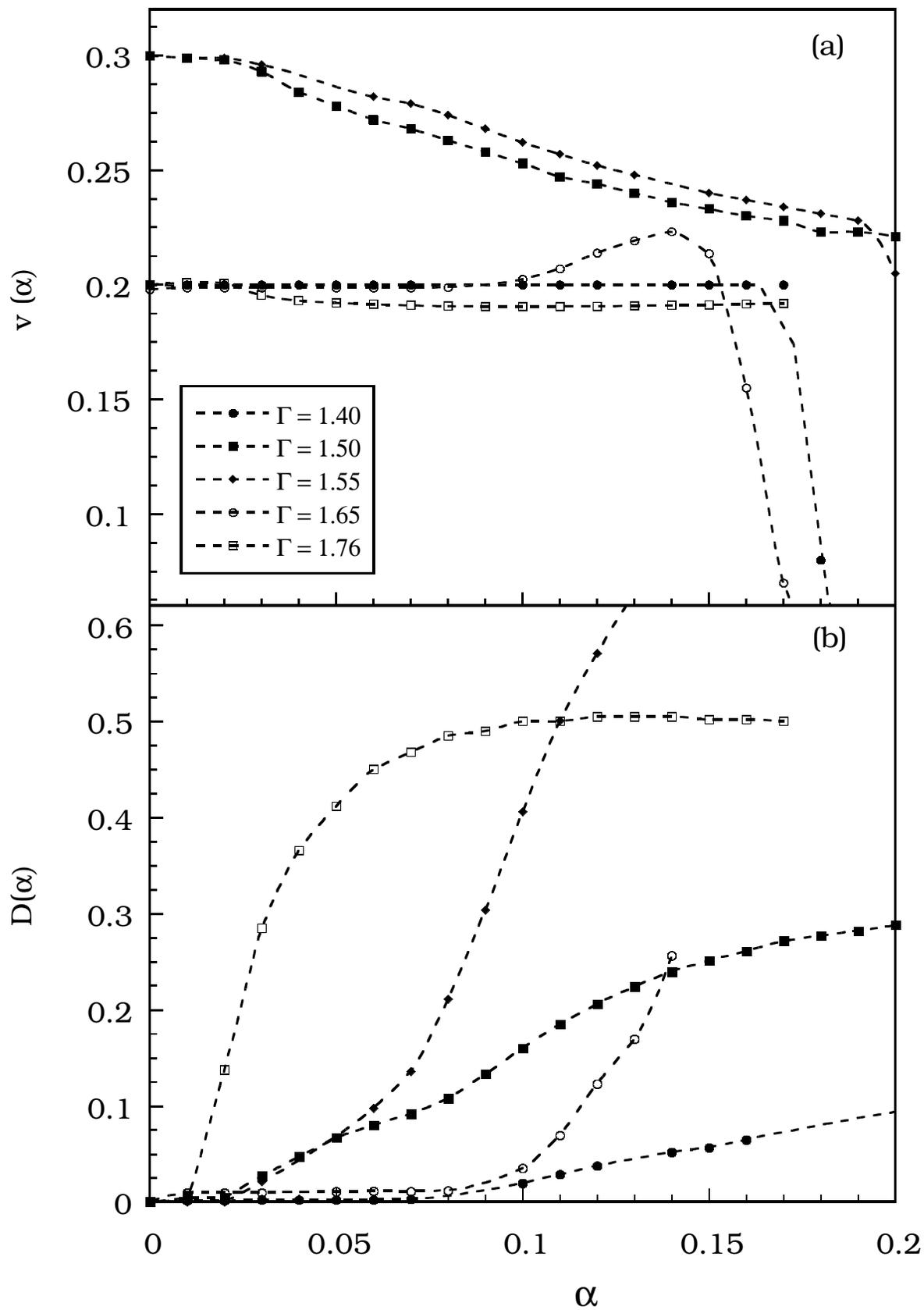
\begin{figure}
\caption{(a) Diffusion coefficient $D(\alpha)$ as a function
of the amount of quenched disorder for
$\Gamma=1.40,~1.50,~1.55,~1.65,~1.76$. (b) Drift velocity
$v(\alpha)$ as a function of the amount of quenched disorder
for $\Gamma=1.40,~1.50,~1.55,~1.65,~1.76$. The parameters are
$\gamma = 1$, $\mu=0.25$, $\omega = 0.1$.}
\label{fig4}
\end{figure}

For several values of the external drive amplitude $\Gamma$,
we have computed from the slopes of the first two moments,
the drift velocity $v(\alpha)$,  and the diffusion
coefficient $D(\alpha)$, as functions of the amount of 
disorder $\alpha$.  The results shown in Fig.\ \ref{fig4}
indicate that  below a ($\Gamma$ dependent) threshold value of
$\alpha$ the drift is slightly  decreasing with increased
quenched disorder, while the diffusion coefficient is steadily
increasing and tends to saturate at high amounts of disorder.
The  fact that disorder has little effect on the drift motion
is explained by the  fact that the drift is a consequence of
the initial asymmetry in the potential, and this asymmetry is
only weakly influenced by small perturbations. We note
here that there is no decrease in the diffusion coefficient
over the  range of disorder considered in this study. This
is in contrast to the decrease of the diffusion coefficient 
observed in other  systems \cite{Marchesoni,Radons}. 
The ``divergence'' of $D(\alpha)$ above the threshold can be 
understood if we consider the fact that $v(\alpha)$ decreases 
to zero. For large enough $\alpha$, some of the trajectories 
in the ensemble become bounded, and their contribution to the 
second moment is proportional to the displacement of 
the center of mass, thus with $t^2$. The number of bounded 
trajectories increases with time, as shown by the steady 
decrease of the drift velocity toward zero. The contribution 
to the second moment (fluctuations) of the $t^2$ term thus 
increases in time, and becomes dominant at late time, 
leading to the above mentioned ``divergence'' of $D(\alpha)$.

There is a number of experimental situations where small 
perturbations of a ratchet potential are relevant, including 
such systems as surface electromigration \cite{barabasi}, 
dielectrophoretic trapping, and particle separation 
techniques \cite{separation}.  
Our preliminary results for the case of a non-negligible 
inertial term in Eq.\ (\ref{motion}) are qualitatively similar 
to the ones for the over-damped case, showing disorder 
induced chaotic diffusion. In this case, however, both the 
``diffusion coefficient'' $D$ and the drift velocity $v$ depend 
on the mass of the particle. Based on the similarities 
mentioned, our results may be relevant for experiments where 
the mass dependence of the drift velocity or diffusion 
coefficient is essential. The efficiency of a nano-scale surface 
smoothening by an ac field suggested by Der\'enyi {\it et al} 
\cite{barabasi} could be actually significantly smaller than 
theoretically predicted because of the chaotic diffusion, induced 
by the inherent "disorder" of a real surface, superimposed on 
the net downhill current. On the other hand, the rough, imperfect 
surface of the electrodes in the dielectrophoretic separation 
technique suggested by  Gorre-Talini {\it et al} \cite{separation} 
can  actually lead to a better efficiency of the process by 
superimposing the chaotic diffusion and drift on top of the 
thermal, Brownian motion. Moreover, the ac-separation techniques 
using a  two-dimensional sieve discussed by Der\'enyi and Astumian 
\cite{separation} can be modified in a very natural way to take 
advantage of the inherent imperfection of the two-dimensional 
structure. This can be done by replacing the Brownian diffusion 
along the drift direction with an additional ac-field along that 
direction. Also, in this way one can fine tune both the drift 
velocity and the diffusion coefficient along the separation 
direction by a convenient choice of the ac-field parameters. 
The temperature can then be used for an independent tuning of the electrophoretic mobility, thus for the transverse displacement.

In summary we have shown that the addition of small amounts 
of quenched disorder in the equation of motion of a continuous 
time system induces a strong diffusive motion. In addition, we 
have found that the presence  of small amounts of disorder 
slightly decreases the regular current (drift motion), but 
significantly increases the transport by chaotic diffusion. 
We have shown also that in the presence of disorder the spatial 
distribution of positions, averaged over the realizations of 
disorder, is described  by a time-dependent Gaussian 
distribution, which is a signature of chaotic motion. These 
unexpected results may help in the interpretation of 
experimental results in studies of friction, particularly at 
the nanoscale, as well as in understanding transport processes 
in molecular motors or designing particle separation techniques.

\bigskip 
\noindent {\bf Acknowledgments}

This work was supported by grants from the Office of Naval 
Research, and from the Universidad Nacional de Mar del Plata. 
A. L. Salas-Brito wants to thank  M.\ Mina and C. Ch. Ujaya 
for their friendly support and acknowledges the partial support 
of CONACyT through grant 1343P-E9607.

\end{document}